\documentclass[12pt]{iopart}%
\usepackage{iopams}         %
\usepackage{amsfonts}       %
\usepackage{amssymb}        %
\usepackage{amsthm}         %

\newcommand{\etat}{\widetilde{\eta}_{\kappa}}        %
\newcommand{\h}{\widetilde{h}^{(\kappa)}}            %
\newcommand{\W}{\widetilde{W}_{\kappa}}              %
\newcommand{\V}{\widetilde{V}_{\kappa}}              %
\newcommand{\xiv}{|\,\xi\rangle_{\kappa}}            %
\newcommand{\zero}{|\,0\rangle_{\kappa}}             %
\newcommand{\D}{\mathcal{D}(\xi_{\kappa})}           %
\newcommand{\bra}{\left[\etat,\etat^{\dagger}\right]}%
\newcommand{\s}{\mathcal{S}_{\kappa}(\xi)}           %
\newcommand{\sk}{\mathcal{S}^{k}_{\kappa}(\xi)}      %
\newcommand{\skk}{\mathcal{S}^{k-1}_{\kappa}(\xi)}   %
\newcommand{\sss}{\mathcal{S}(\xi_{\kappa})}         %
\newcommand{\Q}{\widehat{Q}_{\kappa}}                %
\newcommand{\om}{\omega_\kappa}                      %
\newcommand{\lam}{\lambda_\kappa}                    %
\newcommand{\n}{\nu_\kappa}                          %
\renewcommand{\max}{\textrm{sech}}                   %
\renewcommand{\min}{\textrm{csch}}                   %

\eqnobysec

\begin{document}

\title[Pseudo-Hermitian Coherent states $\ldots$]{Pseudo-Hermitian coherent states under the generalized quantum condition with position-dependent mass}

\author{S A Yahiaoui and M Bentaiba}

\address{LPTHIRM, D\'epartement de physique, Facult\'e des Sciences, Universit\'e Sa\^ad DAHLAB de Blida, B.P. 270 Route de Soum\^aa, 09000 Blida, Algeria}
\ead{s\_yahiaoui@univ-blida.dz and bentaiba@univ-blida.dz}
\begin{abstract}
In the context of the factorization method, we investigate the pseudo-Hermitian coherent states and its Hermitian counterpart coherent states under the generalized quantum condition in the framework of a position-dependent mass. By considering a specific modification in the superpotential, a suitable annihilation and creation operators are constructed in order to reproduce the Hermitian counterpart Hamiltonian in the factorized form. We show that by means of these ladder operators we can construct a wide kind of exactly solvable potentials as well as their accompanying coherent states. Alternatively, we explore the relationship between the pseudo-Hermitian Hamiltonian and its Hermitian counterparts, obtained from a similarity transformation, to construct the associated pseudo-Hermitian coherent states. These latter preserve the structure of Perelomov's states and minimize the generalized position-momentum uncertainty principal.
\end{abstract}

\vspace{2pc}
\submitto{\JPA}
\maketitle

\section{Introduction}%

Motivated by the fact that not only the quantum system with a constant mass but also the quantum system endowed with a position-dependent mass appears as the dynamical algebra, it is of physical and mathematical interest to construct the various coherent states associated with such quantum systems. Recently, the study of the Schr\"{o}dinger equation with the position-dependent mass has attracted a lot of attention \cite{1,2,3,4,5,6,7,8}. This is because such systems have found wide applications in various fields like the study of electronic properties of the semiconductors \cite{9}, $^3$He clusters \cite{10}, quantum wells, wires and dots \cite{11}, quantum liquids \cite{12}, the graded alloys and semiconductor heterostructures \cite{13}, etc.\\
\indent The coherent states have been one of the fastest developing areas in mathematical physics during the last four decades. It was Glauber \cite{14} who showed that the coherent states can be used to describe the electromagnetic correlation functions in the context of quantum optics. As results the construction of such states are defined in three different ways but are all equivalent for the harmonic oscillator: (i) they are eigenstates of annihilation operator \cite{15}, (ii) they minimize the position-momentum uncertainty relation \cite{16} and (iii) they are displaced versions of the ground wave-function \cite{17}. Several approaches and techniques have been used in order to construct the coherent states, namely the Nieto-Simmons method \cite{18}, the irreducible representation of a Lie group \cite{19}, the algebraic method \cite{20}, the supersymmetric quantum mechanics \cite{21,22,23}, and the mixed supersymmetric-algebraic method \cite{24}. Recently, the coherent states endowed with position-dependent mass have been constructed using the intertwining operator \cite{25,26}.\\
\indent On the other hand, there has been a great deal of interest in the properties of pseudo-Hermiticity \cite{27} and $\mathcal{PT}$-symmetric \cite{28} Hamiltonians since it was shown that some of them may have a real and positive spectrum and that these Hamiltonians are prototypes for solvable models of Lie algebraic type \cite{29}. By definition a Hamiltonian $H$ is said to be pseudo-Hermitian if it satisfies $H^{\dag}=\zeta H\zeta^{-1}$, where $\zeta$ is a positive-definite Hermitian operator. However it was shown that such a Hamiltonian is equivalent to a Hermitian counterpart Hamiltonians $h$ according to: $h=\rho H\rho^{-1}$ with $\rho=\sqrt{\zeta}$ \cite{30}.\\
\indent In this context Jones \cite{31} and Bagchi et al. \cite{32} demonstrated that a non-Hermitian Hamiltonian, proposed earlier by Swanson \cite{33},
\begin{eqnarray}\label{1.1}
H^{(\alpha,\beta)}=\omega\left(\eta^{\dagger}\eta+\frac{1}{2}\right)+\alpha\,\eta^{2}
+\beta\,\eta^{\dagger 2},
\end{eqnarray}
with $\omega$, $\alpha$ and $\beta$ are three real parameters, admits an equivalent Hermitian Hamiltonian $h^{(\alpha,\beta)}$ where $\eta$ and $\eta^{\dagger}$ obeying the standard commutation relation; i.e., $\left[\eta,\eta^{\dagger}\right]=1$.\\
\indent The aim of this paper is to construct formally a set of position-dependent mass coherent states for Hermitian counterpart Hamiltonians (HCS) $\widetilde{h}^{(\kappa)}$ under a generalized quantum condition $\bra=(\kappa+1)F[x]$, where $F[x]$ is some generating functional. Afterward we explore a similarity transformation, which maps the original Hamiltonian $\widetilde{H}^{(\kappa)}$ onto $\widetilde{h}^{(\kappa)}$, to deduce the associated pseudo-Hermitian coherent states (PHCS). Indeed the ladder operators $\eta$ and $\eta^{\dagger}$ in \eref{1.1} based upon the deformed momentum operator $\Pi=U(x)\,p\,U(x)$ lead to a corresponding Hermitian counterpart Hamiltonian with a coordinate dependence in mass, however, it does not reproduce the position-dependent mass in the factorized form. In other words, they do not act as ladder operators on the eigenfunctions. Fortunately, we will see that this inadequacy of the usual approach may be circumvented by invoking a modification in the superpotential provided with the constraint $\beta=0$ (or equivalently $\alpha=0$ due to the symmetric nature of new Hermitian Hamiltonian $\h$, where $\kappa$ is either $\alpha$ or $\beta$.) Then the new ladder operator $\etat$ and $\etat^{\dagger}$ are nothing but annihilation and creation operators endowed within the generally deformed oscillator algebra. The corresponding coherent states are then constructed with the modified superpotential under the generalized quantum condition. It is found that the analytical expressions of such HCS (i.e., PHCS) preserve the structure of Perelomov's approach and minimize the generalized position-momentum uncertainty principal. The method of solution is applied to a wide kind of exactly solvable potentials and the corresponding HCS are constructed.\\
\indent We organized our paper as follows. In section 2 we apply the concept of the pseudo-Hermiticity to the Hamiltonian \eref{1.1} deducing the corresponding Hermitian Hamiltonian $h^{(\alpha,\beta)}$. A new Hermitian counterpart Hamiltonian $\h$ is reproduced in the factorized form through the modified creation and annihilation operators. In section 3, the HCS are constructed under a generalized quantum condition in the context of Perelomov's states and exploring the relationship between the pseudo-Hermitian Hamiltonians and its Hermitian counterpart, obtained from the similarity transformation, a set of PHCS is constructed straightforwardly. Applying the procedure of the section 3 various exactly solvable potentials as well as their HCS are constructed in section 4. Finally, we present our conclusions in the last section.

\section{Modified hermitian counterpart within generalized quantum condition}%

In units wherein $\hbar=m_0=1$, the general first-order differential form for $\eta$ and $\eta^{\dagger}$ are given by \cite{25}
\begin{equation}\label{2.1}
\eqalign{\eta&=\frac{1}{\sqrt{2}}\left(U(x)\,\frac{\rmd}{\rmd x}\,U(x)+W(x)\right),\cr
\eta^{\dagger}&=\frac{1}{\sqrt{2}}\left(-U(x)\,\frac{\rmd}{\rmd x}\,U(x)+W(x)\right),}
\end{equation}
with $U(x)$ and $W(x)$ are real functions. Here the generalized quantum condition yields
\begin{equation}\label{2.2}
\left[\eta,\eta^{\dagger}\right]=U^2(x)W'(x),
\end{equation}
where the prime denotes derivative with respect to $x$. Substituting \eref{2.1} into the Hamiltonian \eref{1.1}, the corresponding eigenvalues equation reads as
\begin{eqnarray}\label{2.3}
\left(-\frac{\Omega^{(-)}}{2}\,U^4(x)\,\frac{\rmd^2}{\rmd x^2}+K_{\alpha,\beta}(x)\,\frac{\rmd}{\rmd x}+R_{\alpha,\beta}(x)\right)\psi(x)=E\,\psi(x),
\end{eqnarray}
with
\begin{eqnarray}\label{2.4}
K_{\alpha,\beta}(x)&=(\alpha-\beta)\,U^2(x)W(x)-2\,\Omega^{(-)}U^3(x)\,U'(x)\\
R_{\alpha,\beta}(x)&=\frac{1}{2}\,\Big\{\omega\left[1-U^2(x)W'(x)\right]+\Omega^{(+)}W^2(x)
+(\alpha-\beta)\left[U^2(x)W(x)\right]'\nonumber\\
&-\Omega^{(-)}U^2(x)\left[2\,U'^2(x)+U(x)\,U''(x)\right]\Big\}
\end{eqnarray}
and $\Omega^{(\pm)}=\omega\pm\alpha\pm\beta$.\\
\indent In order to bring \eref{2.3} to a Schr\"{o}dinger equation in the Hermitian form, it is straightforward to remove the first-derivative term in \eref{2.3}. For this purpose defining a similarity transformation $\psi(x)=\rho_{\alpha,\beta}^{-1}(x)\chi(x)$ \cite{32}, where $\rho_{\alpha,\beta}(x)$ is defined as
\begin{eqnarray}\label{2.6}
\rho_{\alpha,\beta}(x)&=A^s(x)\exp\left[-\int^{x}\frac{K_{\alpha,\beta}(y)}{U^4(y)}\,\rmd y\right]\nonumber\\
&=A^s(x)\,\left[U(x)\right]^{2\Omega^{(-)}}\exp\left[-(\alpha-\beta)\int^{x}\frac{W(y)}{U^2(y)}\,\rmd y\right],
\end{eqnarray}
with $A(x)$ is an unknown function and $s$ a parameter to be determined. Without loss of generality, we set $\omega=\alpha+\beta+1$, i.e., $\Omega^{(-)}=1$. To determine $A(x)$ and $s$ we substitute \eref{2.4} and \eref{2.6} into \eref{2.3} we get $A(x)=U(x)$ and $s=-2$, then a Hermitian Hamiltonian $h^{(\alpha,\beta)}=\rho_{\alpha,\beta}(x)H^{(\alpha,\beta)}\rho_{\alpha,\beta}^{-1}(x)$ is equivalent to $H^{(\alpha,\beta)}$. Therefore $\rho_{\alpha,\beta}(x)$, $h^{(\alpha,\beta)}$ and the associated effective potential $V_{\rm eff}^{(\alpha,\beta)}(x)$ read explicitly as
\begin{equation}\label{2.7}
\eqalign{\rho_{\alpha,\beta}(x)&=\exp\left[-(\alpha-\beta)\int^{x}\frac{W(y)}{U^2(y)}\,\rmd y\right],\cr
h^{(\alpha,\beta)}&=-\frac{1}{2}\frac{\rmd}{\rmd x}\,U^4(x)\,\frac{\rmd}{\rmd x}+V_{\rm eff}^{(\alpha,\beta)}(x),\cr
V_{\rm eff}^{(\alpha,\beta)}(x)&=\frac{\omega^2-4\alpha\beta}{2}\,W^2(x)-\frac{\omega}{2}\,U^2(x)W'(x)
+\frac{\omega}{2}+\mathcal{V}_{U}(x),}
\end{equation}
where $\mathcal{V}_{U}(x)=-\,U^2(x)\,U'^{2}(x)-U^3(x)\,U''(x)/2$ is a depending dimensionless mass potential.\\
\indent We need to notice that $h^{(\alpha,\beta)}$ is symmetric with respect to the parameters $\alpha$ and $\beta$, and the condition $\rho_{\alpha,\beta}(x)=\rho_{\beta,\alpha}^{-1}(x)$ holds \cite{32}, so that ${H^{(\alpha,\beta)}}^{\dagger}=\zeta_{\alpha,\beta}H^{(\alpha,\beta)}\zeta_{\alpha,\beta}^{-1}$ is pseudo-Hermitian, with
\begin{eqnarray}\label{2.8}
\zeta_{\alpha,\beta}&\equiv &\rho_{\beta,\alpha}^{-1}(x)\rho_{\alpha,\beta}(x)\nonumber\\
&=&\exp\left[-2\,(\alpha-\beta)\int^{x}\frac{W(y)}{U^2(y)}\,\rmd y\right]>0,
\end{eqnarray}
for $U(x)>0$. The second equation in \eref{2.7} clearly reveals that for coordinate dependent mass, $U^4(x)$ could be identified with the inverse of certain mass function, i.e., $U(x)=m^{-1/4}(x)$, with $M(x)=m_{0}m(x)$ and $m(x)$ is a dimensionless mass. A consequence of this is that $h^{(\alpha,\beta)}$ is not factorized under the ladder operators $\eta$ and $\eta^{\dagger}$ because they do not give a closed algebra.\\
\indent It should be pointed out that it is difficult to obtain the suitable ladder operators as defined in conventional quantum mechanics which permit the factorization of $h^{(\alpha,\beta)}$ in \eref{2.7}. The idea is to develop the scheme of a factorization method by reducing the number of parameters $\alpha$ and $\beta$ to one real parameter $\kappa$ by making use of the modified superpotential, so that the new annihilation operator, $\etat$, annihilates the ground-state.\\
\indent To this end we can recast $h^{(\alpha,\beta)}$ in the new form
\begin{eqnarray}\label{2.9}
h^{(\alpha,\beta)}\rightarrow\h=\etat^{\dagger}\etat+\delta_{\kappa},
\end{eqnarray}
where $\kappa$ is either $\alpha$ or $\beta$, while $\etat$ and $\etat^{\dagger}$ are expressed in terms of the modified superpotential $\W(x)=p\,\mathcal{W}(x)+q\,U'(x)$ as
\begin{equation}\label{2.10}
\eqalign{\etat&=\frac{1}{\sqrt{2}}\left(U(x)\,\frac{\rmd}{\rmd x}\,U(x)+\W(x)\right),\cr
\etat^{\dagger}&=\frac{1}{\sqrt{2}}\left(-U(x)\,\frac{\rmd}{\rmd x}\,U(x)+\W(x)\right),}
\end{equation}
with $\mathcal{W}(x)=W(x)+q_{1}U'(x)$ and $p$, $q$ and $q_{1}$ are some real constants depending on $\kappa$.\\
\indent Substituting \eref{2.10} into \eref{2.9} and comparing with \eref{2.7} we get two equivalent solutions with respect of parameters $\alpha$ and $\beta$, as well as a constraint controlling the parameters $p$, $q$ and $q_{1}$
\begin{equation}\label{2.11}
p=\cases{\alpha+1&for $\beta=0$\\
\beta+1&for $\alpha=0$\\},\qquad \textrm{and}\qquad p\,q_{1}+q=0.
\end{equation}

\indent As a consequence it is found that the modified superpotential is $\W(x)=(\kappa+1)W(x)$, no matter what the explicit form of $q$ and $q_{1}$ is. Then the general form of the potential $\V(x)$ is given by
\begin{eqnarray}\label{2.12}
\V(x)=\frac{\left(\kappa+1\right)^2}{2}\,W^2(x)-\frac{\kappa+1}{2}\,U^2(x)W'(x)+\delta_\kappa,
\end{eqnarray}
where $\delta_{\kappa}=(\kappa+1)/2$. Then it is easy to verify that $\etat$, $\etat^{\dagger}$ and the Hamiltonian $\h$ satisfy the mutually commutation relations
\begin{equation}\label{2.13}
\eqalign{\bra&=(\kappa+1)\,U^2(x)W'(x),\cr
\left[\h,\etat^{\dagger}\right]&=(\kappa+1)\,\etat^{\dagger}\,U^2(x)W'(x),\cr
\left[\h,\etat\right]&=-(\kappa+1)\,U^2(x)W'(x)\,\etat.}
\end{equation}

\indent This algebra is nothing but the generally deformed oscillator algebra with the structure function $(\kappa+1)\:U^2(x)W'(x)$. And as a consequence the simplest form of a (shifted) harmonic oscillator can be taken for $\bra=\kappa+1$ which leads to express $W(x)=\mu(x)+\rm const.$, where hereafter we adopt the definition $\mu(x)=\int^{x}\rmd y/U^2(y)$.\\
\indent Now by acting \eref{2.9} on the state $|v\rangle$ belonging on the Hilbert space $\mathcal{H}$, one gets the Hamiltonian in the factorized form
\begin{equation}\label{2.14}
\etat^{\dagger}\etat|v\rangle=\Delta E^{(\kappa)}_{0,v}|v\rangle,
\end{equation}
where $\Delta E^{(\kappa)}_{0,v}=E_v-\delta_{\kappa}$.\\
\indent It is worth noting that the energy eigenvalues expressed in \eref{2.14} are chosen such that $\delta_{\kappa}$ corresponds to the ground-state energy $\delta_{\kappa}=\epsilon_{0,\kappa}$. Then, if $\epsilon_{0,\kappa}\neq 0$ (i.e., $\kappa\neq -1$ which is as it should be since the potential in \eref{2.12} vanishes) then, due to \eref{2.13}, $\etat$ and $\etat^{\dagger}$ are the ladder operators of the position-dependent mass system.

\section{Construction of Hermitian and pseudo-Hermitian coherent states}%

Our key aim now is to construct a set of position-dependent mass Hermitian coherent states (HCS) $\xiv$ for the Hamiltonian $\h$, and by exploring that there exists a similarity transformation which maps the pseudo-Hermitian Hamiltonian adjointly to a Hermitian Hamiltonian, we deduce the PHCS $|\,\Xi\rangle_\kappa$ with regard to the properties of the metric $\rho_\kappa^{-1}(x)$.\\
\indent Naturally, we expect to find many shapes for $|\,\Xi\rangle_\kappa$ due to the feature that the similarity transformation is not unique when the only requirement for $\h$ is its Hermiticity \cite{29}. However, the determination of PHCS may be achieved by assuming some conditions that are brought on functions associated with coherent states.

\subsection{Position dependent-mass Hermitian coherent states}%

A position-dependent mass is shown to be factorized by two adjoint operators $\etat$ and $\etat^{\dagger}$, then the related coherent states $\xiv$ for the Hamiltonian $\h$ (i.e., for the potential \eref{2.12}) are constructed as eigenstates of the annihilation operator $\etat\xiv=\xi_{\kappa}\xiv$ and $\etat$ annihilates the ground-state $\etat\zero=0$.\\
\indent The ground-state $\zero$ can be calculated by integrating the first equation of \eref{2.10}
\begin{equation}\label{3.1}
\zero=U^{-1}(x)\exp\left[-(\kappa+1)\int^{\mu(x)}W(y)\,\rmd\mu(y)\right],
\end{equation}
where we have used the definition of $\mu(x)$ quoted above. Here the presence of $\kappa$ in our previous results suggests that the complex parameter $\xi_{\kappa}$ depends on some real parametric function $\gamma(\kappa)$ such as $\xi_{\kappa}=\gamma(\kappa)\,\xi$. The choice of the function $\gamma(\kappa)$ in the sense of this paper corresponds to fixing the form of  HCS and PHCS as reviewed below.\\
\indent As usual, the coherent states can be frequently introduced by using the displacement and unitary operator
\begin{equation}\label{3.2}
\xiv=\D\zero,
\end{equation}

\indent To this purpose, let us assume that $\D$ acts on the ladder operators $\etat$ and $\etat^{\dagger}$ according to the generalized scheme
\begin{equation}\label{3.3}
\eqalign{\D^{\dagger}\,\etat\,\D&=\etat+\xi_{\kappa}\bra,\cr
\D^{\dagger}\,\etat^{\dagger}\,\D&=\etat^{\dagger}+\xi_{\kappa}^{\ast}\bra.}
\end{equation}

\indent The presence of the commutation relations in \eref{3.3} is due to the fact that the generalized quantum condition \eref{2.13} is considered which, as we are going to show, is able to construct a suitable set of HCS. It is clear that if $\bra=\rm const.$, then we obtain the usual (shifted) harmonic oscillator coherent states.\\
\indent In order to construct the coherent states $\xiv$ of \eref{3.2}, we look for $\D$ as \cite{25}
\begin{equation}\label{3.4}
\D=\exp\left[\rmi\,\sss\right],
\end{equation}
where $\sss\equiv\s=\gamma(\kappa)\,\mathcal{S}(\xi)$ is a real and linear function on $\xi_{\kappa}$ and verifies the relation $\s=-\rmi\,\xi_{\kappa}\Q$, where $\Q$ is an operator to be determined. Moreover as $\D$ is an unitary one leads to identify that both $\s$ and $\Q$ are Hermitians if $\xi_{\kappa}=-\xi^{\ast}_{\kappa}$, i.e., $\Re(\xi_{\kappa})=0$. Indeed by imposing the requirements
\begin{equation}\label{3.5}
\eqalign{\left[\s,\etat\right]&=\rmi\,\xi_{\kappa}\bra,\cr
\left[\s,\etat^{\dagger}\right]&=-\rmi\,\xi_{\kappa}\bra,}
\end{equation}
and substituting $\s=-\rmi\,\xi_{\kappa}\Q$ into \eref{3.5} a brief examination yields two solutions for $\Q$; either $\etat^{\dagger}$ or $\etat+\etat^{\dagger}$. It is obvious that the second solution is that which we are interested because the first solution will be omitted in order to avoid ill-defined Hermiticity condition imposed to $\Q$, while the second one combined with the condition $\xi_{\kappa}=-\xi^{\ast}_{\kappa}$ yields to the well-known deformed displacement operator $\exp\left[\xi_{\kappa}\etat^{\dagger}-\xi^{\ast}_{\kappa}\etat\right]$. At this stage, it is important to stress that \eref{3.5} verify \eref{3.3}, respectively.

\begin{proof}
Let us consider the operator $\widehat{\mathcal{P}}_{\kappa}=\left[\etat,\D\right]$. Expanding $\exp\left[\rmi\,\s\right]$ in the development of the Taylor series
\begin{equation}\label{3.6}
\widehat{\mathcal{P}}_{\kappa}=-\sum_{k=0}^{\infty}\frac{\rmi^k}{k!}\left[\sk,\etat\right],
\end{equation}
and using \eref{3.5} the straightforward calculation leads to a recursion relation which is satisfied by the commutators
\begin{equation}\label{3.7}
\left[\sk,\etat\right]=\rmi\,k\,\xi_{\kappa}\,\skk\bra,
\end{equation}
and inserting \eref{3.7} into \eref{3.6} we have
\begin{equation}\label{3.8}
\widehat{\mathcal{P}}_{\kappa}=\xi_{\kappa}\,\rme^{\rmi\,\s}\bra.
\end{equation}

\indent On the other hand starting from $\widehat{\mathcal{P}}_{\kappa}=\left[\etat,\D\right]$ and multiplying both sides of the expression on the left by $\D^{\dagger}$ and comparing the result with \eref{3.3} we obtain \eref{3.8}. Then the calculations performed agree with our assertion.
\end{proof}

\indent Hence substituting the expressions of $\s$ and $\Q$ into \eref{3.4} and this latter into \eref{3.2} including \eref{3.1}, the HCS for the potential \eref{2.12} can be specified by the general formula
\begin{eqnarray}\label{3.9}
\xiv &=\D\zero\nonumber\\
&=m^{1/4}(x)\,\rme^{\sqrt{2}\,(\kappa+1)\gamma(\kappa)\,\xi\,W(x)}\exp\left[-(\kappa+1)
\int^{\mu(x)}W(y)\,\rmd\mu(y)\right],
\end{eqnarray}
where $\Q\equiv\etat+\etat^{\dagger}=\sqrt{2}\,(\kappa+1)W(x)$ is deduced from \eref{2.10} and for coordinate dependent mass we have $U^{-1}(x)= m^{1/4}(x)$.\\
\indent Using the condition $\xi_{\kappa}=-\xi_{\kappa}^{\ast}$, a simple inspection of \eref{3.9} shows that $_{\kappa}\langle\xi\xiv=\,_{\kappa}\langle 0\zero=1$ and, following \cite{24,25}, it is easy to verify that these states minimize the generalized position-momentum uncertainty relation which yields
\begin{eqnarray}\label{3.10}
\langle\Delta\W\rangle^2\langle\Delta\Pi_\kappa\rangle^2&=\frac{1}{4}\,
_{\kappa}\langle\xi|\bra\xiv^2\nonumber\\
&=\frac{\left(\kappa+1\right)^2}{4}\,
_{\kappa}\langle\xi|U^2(x)W'(x)\xiv^2,
\end{eqnarray}
where $\Pi_\kappa$ is the deformed momentum operator defined as $\Pi_\kappa\equiv-\rmi(\etat-\etat^{\dagger})/\sqrt{2}=U(x)p\,U(x)$ where $p=-\rmi\,\rmd/\rmd x$.

\subsection{Position-dependent mass Pseudo-Hermitian coherent states}%

As it has been mentioned in Ref. \cite{32}, the associated non-Hermitian Hamiltonian $\widetilde{H}^{(\alpha,\beta)}$ has the same real spectrum with the ground-state $|\,\widetilde 0\rangle=\rho^{-1}_{\alpha,\beta}(x)|\,0\rangle.$\\
\indent As it should be expected we are thus in a position to construct PHCS, which to this order is given by a similarity transformation
\begin{eqnarray}\label{4.1}
|\,\Xi\rangle=\rho^{-1}_{\alpha,\beta}(x)|\,\xi\rangle.
\end{eqnarray}
\indent However, the requirement that the only knowledge for $\h$ is its Hermiticity leads to suggest that a similarity transformation is not unique \cite{29}; on other words the determination of PHCS can be achieved using some restrictions imposed to the metric $\rho_{\alpha,\beta}^{-1}(x)$. To this end, it is worth mentioning the existence of an underlying indicial symmetry that explain how a sufficient condition for the positivity of $\zeta_{\alpha,\beta}$ may be provided by interchanging $\alpha$ to $\beta$ and vice versa. This is clearly taking into account that $\rho_{\alpha,\beta}(x)=\rho^{-1}_{\beta,\alpha}(x)$, so that the following symmetry in terms of $\kappa$ is kept
\begin{eqnarray}\label{4.2}
\rho_{-\kappa}(x)=\rho^{-1}_{\kappa}(x).
\end{eqnarray}
\indent The most straightforward assumption to consider for $\rho_\kappa(x)$ is an exponential form as the one chosen in \eref{3.9}
\begin{eqnarray}\label{4.3}
\rho_{\kappa}(x)=\exp\left[-f(\kappa)\int^{\mu(x)} W(y)\:\rmd\mu(y)\right],
\end{eqnarray}
where $f(\kappa)$ is some unknown function to be determined. Besides the restriction \eref{4.3}, the condition \eref{4.2} leads to identify $f(\kappa)$ as an odd-function, i.e., $f(-\kappa)=-f(\kappa)$. Then inserting \eref{3.9} into \eref{4.1}, PHCS $|\Xi\rangle_\kappa$ are re-expressed as
\begin{eqnarray}\label{4.4}
|\Xi\rangle_\kappa&=&\rho^{-1}_{\kappa}(x)|\xi\rangle_\kappa\nonumber\\
&=&m^{1/4}(x)\,\rme^{\sqrt{2}\,(\kappa+1)\gamma(\kappa)\xi\,W(x)}\exp\left[-(\kappa+1-f(\kappa))
\int^{\mu(x)}W(y)\,\rmd\mu(y)\right].
\end{eqnarray}

\indent The PHCS \eref{4.4} are ideally suited with regard to the determination of $f(\kappa)$. To start with we impose on $|\,\Xi\rangle_\kappa$ by demanding it to be on the form of $|\,\xi\rangle_\kappa$. This means we may assume the equality
\begin{eqnarray}\label{4.5}
(\kappa+1)\gamma(\kappa)\equiv\kappa+1-f(\kappa),
\end{eqnarray}
which is enough to determine both $\gamma(\kappa)$ and $f(\kappa)$. Note that when we impose the constraint $f(-\kappa)=-f(\kappa)$ to the equality \eref{4.5} we find that this latter is solved solely by demanding to $\gamma(\kappa)$ to be an odd-function too\footnote{Indeed if $\gamma(\kappa)$ is an even-function then $f(\kappa)=0$, which we will neglect.}, i.e.,
\begin{eqnarray}\label{4.6}
\gamma(\kappa)=\frac{1}{\kappa}\qquad\Rightarrow\qquad f(\kappa)=\kappa-\frac{1}{\kappa},\qquad \textrm{for}\quad\kappa\neq 0,\pm 1.
\end{eqnarray}
\indent Thus by making use of the constraining equation \eref{4.6}, we can re-express HCS $|\,\xi\rangle_\kappa$ in \eref{3.9} and PHCS $|\,\Xi\rangle_\kappa$ in \eref{4.4} purely as a function of $\overline{W}_{\kappa}(x)$
\begin{eqnarray}\label{4.7}
|\xi\rangle_\kappa&\equiv m^{1/4}(x)\,\rme^{\sqrt{2}\,\xi\,\overline{W}_{\kappa}(x)}\exp\left[-\kappa
\int^{\mu(x)}\overline{W}_{\kappa}(y)\,\rmd\mu(y)\right],
\end{eqnarray}
and
\begin{eqnarray}\label{4.8}
|\Xi\rangle_\kappa&\equiv m^{1/4}(x)\,\rme^{\sqrt{2}\,\xi\,\overline{W}_{\kappa}(x)}\exp\left[-
\int^{\mu(x)}\overline{W}_{\kappa}(y)\,\rmd\mu(y)\right],
\end{eqnarray}
where
\begin{eqnarray}\label{4.9} \overline{W}_{\kappa}(x)\equiv\frac{1}{\kappa}\,\widetilde{W}_{\kappa}(x)=\frac{\kappa+1}{\kappa}\,W(x).
\end{eqnarray}

\indent The identities \eref{4.7} and \eref{4.8} are our main results.

\section{An exactly solvable potentials and their Hermitian coherent states}%

Here we illustrate the procedure by which a wide kind of exactly solvable potentials endowed with position-dependent mass can be recovered as well as their corresponding coherent states. Its forms must satisfy the commutation relation \eref{2.13} which the explicit expression $U^2(x)W'(x)$ is equal to a certain generating functional $F[x]$ to be determined.\\
\indent A deeper insight is necessary if we are interested to find solution of \eref{2.13}. And for convenience we introduced the auxiliary function $\varphi(x)$ governed by some differential equations and related to $W(x)$. The strategy consists in choosing these differential equations in such a way that $U^2(x)\varphi'(x)$ contains some terms which allows us to get rid derivatives and, as a consequence, the solutions in $\varphi(x)$; i.e., $W(x)$, can be explicitly carried out by a simple Euler's type integration.\\
\indent Our solutions fall into three classes and differ slightly from those proposed in \cite{2}\footnote{Comparing our formulas \eref{5.1}-\eref{5.3} with (3.5)-(3.16) quoted in Ref. \cite{2}.} with which we can obtain the Hermitian potentials, the ground-state energy eigenvalue and the accompanying HCS. They are identified with differential equations
\begin{enumerate}
\item Class 1:
  \begin{equation}\label{5.1}
  \eqalign{W(x)=k_0\,\varphi(x)+k_1,\cr
  U^2(x)\varphi'(x)=a\,\varphi^2(x)+b\,\varphi(x)+c,}
  \end{equation}
\item Class 2:
  \begin{equation}\label{5.2}
  \eqalign{W(x)=k_0\,\varphi(x)+\frac{k_1}{\varphi(x)},\cr
  U^2(x)\varphi'(x)=a\,\varphi^2(x)+b,}
  \end{equation}
\item Class 3:
  \begin{equation}\label{5.3}
  \eqalign{W(x)=\frac{k_0\,\varphi(x)+k_1}{\sqrt{a^2\,\varphi^2(x)+b^2}},\cr U^2(x)\varphi'(x)=(c\,\varphi(x)+d)\sqrt{a^2\,\varphi^2(x)+b^2},}
  \end{equation}
\end{enumerate}
where $\mathbf{k}=(k_0,k_1)$ and $\mathbf{a}=(a,b,c,d)$ are two sets of parameters which determined completely the quantum system.\\
\indent As we can see, and for lack of space, we have not exhibited the detailed results of our calculations which can be easily determined from \eref{2.12} and \eref{5.1}-\eref{5.3}. The corresponding HCS are deduced for the Hermitian counterpart Hamiltonian using \eref{4.7} while \eref{4.8} is used to convert them to PHCS.

\paragraph{Shifted harmonic oscillator (class 1): $(a=b=0,\,c=1)$}%
\begin{eqnarray}
\fl\varphi(x)&=\mu(x),\qquad\qquad W(x)=k_0\mu(x)+k_1,\nonumber\\
\fl\V(x)&=\frac{\om^2}{2}\left(\mu(x)-\frac{\lam}{\om}\right)^2,\qquad\qquad
\epsilon_{0,\kappa}=\frac{\om}{2},\nonumber\\
\fl\xiv&=m^{1/4}(x)\,\exp\Big\{\sqrt{2}\,\left(\om\mu(x)-\lam\right)\xi_\kappa\Big\}
\exp\Big\{-\frac{\om}{2}\,\mu^2(x)+\lam\mu(x)\Big\},\label{5.4}
\end{eqnarray}
where $\omega_{\kappa}=(\kappa+1)k_0$ and $\lambda_{\kappa}=-(\kappa+1)k_1$. The HCS of \eref{5.4} are eigenstates of $\etat$ which minimize the uncertainty relation \eref{2.13} for $\bra= \omega_{\kappa}=\rm const.$

\paragraph{Morse potential (class 1): $(a=0,\,c=-b,\,k_1=0)$}%
\begin{eqnarray}
\fl\varphi(x)&=1-\frac{1}{c}\,\rme^{-c\mu(x)},\qquad\qquad W(x)=k_0-\frac{k_0}{c}\,\rme^{-c\mu(x)},\nonumber\\
\fl\V(x)&=\frac{\lam^2 c^2}{2}\,\rme^{-2c\mu(x)}+j_\kappa\lam\,\rme^{-c\mu(x)},\qquad\qquad
\epsilon_{0,\kappa}=-\frac{c^2}{8}\left(2j_\kappa+1\right)^2,\nonumber\\
\fl\xiv&=m^{1/4}(x)\exp\Bigg\{\sqrt{2}\left[\lam c^2+\left(j_\kappa+\frac{1}{2}\right)\,\rme^{-c\mu(x)}\right]\xi_\kappa\Bigg\}\,\rme^{-\lam c^2\mu(x)}\nonumber\\
\fl &\times\exp\big\{-\lam\,\rme^{-c\mu(x)}\big\},\label{5.5}
\end{eqnarray}
where $\lam=(\kappa+1)k_0/c^2$ and $j_\kappa=-(\lam c+1/2)$. The HCS of \eref{5.5} minimize the uncertainty relation $\bra=\lam c^2\,\rme^{-c\mu(x)}$.

\paragraph{Radial Coulomb potential (class 1): $(b^2=4ac,\,k_1=0)$}%
\begin{eqnarray}
\fl\varphi(x)&=-\frac{1}{a\mu(x)}-\frac{b}{2a},\qquad\qquad W(x)=-\frac{k_0}{a\mu(x)}-\frac{bk_0}{2a},\nonumber\\
\fl\V(x)&=-\frac{Ze^2}{\mu(x)}+\frac{l_\kappa(l_\kappa+1)}{2\mu^2(x)},\qquad\qquad
\epsilon_{0,\kappa}=-\frac{1}{2}\left(\frac{Ze^2}{l_\kappa+1}\right)^2,\nonumber\\
\fl\xiv&=m^{1/4}(x)\exp\Bigg\{\sqrt{2}\left[-\frac{l_\kappa+1}{\mu(x)}+\frac{Ze^2}{l_\kappa+1}\right]
\xi_\kappa\Bigg\}\left[\mu(x)\right]^{l_\kappa+1}\exp\Bigg\{{-\frac{Ze^2}{l_\kappa+1}\mu(x)}\Bigg\}
,\label{5.6}
\end{eqnarray}
where $l_\kappa=(\kappa+1)k_0/a-1$ and $Ze^2=-b(l_\kappa+1)^2/2$. The HCS of \eref{5.6} are eigenstates of $\etat$ which minimize the uncertainty relation for $\bra=(l_\kappa+1)/\mu^2(x)$.

\paragraph{P\"oschl-Teller potential (class 1): $(a=-c,\,b=k_1=0)$}%
\begin{eqnarray}
\fl\varphi(x)&=\tanh{a\mu(x)},\qquad\qquad W(x)=k_0\tanh{a\mu(x)},\nonumber\\
\fl\V(x)&=-a^2\frac{j_\kappa(j_\kappa+1)}{2}\,\max^{2}a\mu(x),\qquad\qquad
\epsilon_{0,\kappa}=-\frac{a^2}{2}j_\kappa^2,\nonumber\\
\fl\xiv&=m^{1/4}(x)\exp\Big\{\sqrt{2}j_\kappa a\,\xi_\kappa\tanh a\mu(x)\Big\}\left[\cosh a\mu(x)\right]^{-j_\kappa},\label{5.7}
\end{eqnarray}
where $j_\kappa=(\kappa+1)k_0/a$. The HCS of \eref{5.7} minimize the uncertainty relation $\bra=j_\kappa a^2\,\max^2 a\mu(x)$.

\paragraph{Eckart potential (class 1): $(a=c,\,b=0)$}%
\begin{eqnarray}
\fl\varphi(x)&=\coth{a\mu(x)},\qquad\qquad W(x)=-k_0\coth{a\mu(x)}+k_1,\nonumber\\
\fl\V(x)&=\frac{a^2}{2}\lam(\lam-1)\,\min^{2}a\mu(x)-\n a^2\coth a\mu(x),\qquad\qquad
\epsilon_{0,\kappa}=-\frac{a^2}{2}\left(\lam^2+\frac{\n^2}{\lam^2}\right),\nonumber\\
\fl\xiv&=m^{1/4}(x)\exp\Bigg\{\sqrt{2}\,a\left[-\lam\coth a\mu(x)+\frac{\n}{\lam}\right]\xi_\kappa\Bigg\}\exp\Bigg\{-\frac{a\n}{\lam}\mu(x)\Bigg\}\nonumber\\
\fl &\times\left[\sinh a\mu(x)\right]^{\lam},\label{5.8}
\end{eqnarray}
where $\lam=(\kappa+1)k_0/a$ and $a^2\n=(\kappa+1)^{2}k_0k_1$. The HCS of \eref{5.8} are eigenstates of $\etat$ which minimize the uncertainty relation for $\bra=\lam a^2\,\min^2 a\mu(x)$.

\paragraph{Rosen-Morse potential (class 1):}%
\begin{eqnarray}
\fl\varphi(x)&=\cot{a\mu(x)},\quad\qquad W(x)=k_0\cot{a\mu(x)}-k_1,\nonumber\\
\fl\V(x)&=\frac{a^2}{2}\lam(\lam+1)\,\csc^{2}a\mu(x)-\n a^2\cot a\mu(x),\quad\qquad \epsilon_{0,\kappa}=\frac{a^2}{2}\left(\lam^2-\frac{\n^2}{\lam^2}\right),\nonumber\\
\fl\xiv&=m^{1/4}(x)\exp\Bigg\{\sqrt{2}\,a\left[\lam\cot a\mu(x)-\frac{\n}{\lam}\right]\xi_\kappa\Bigg\}\exp\Bigg\{\frac{a\n}{\lam}\mu(x)\Bigg\}\left[\sin a\mu(x)\right]^{-\lam},\label{5.9}
\end{eqnarray}
where $\lam=(\kappa+1)k_0/a$ and $a^2\n=(\kappa+1)^{2}k_0k_1$. The HCS of \eref{5.9} are eigenstates of $\etat$ which minimize the uncertainty relation for $\bra=-\lam a^2\csc^2 a\mu(x)$.

\paragraph{Manning-Rosen potential (class 1): $(a=-1,\,b>0,\,c=0)$}%
\begin{eqnarray}
\fl\varphi(x)&=\frac{b\,\rme^{-b\mu(x)}}{1-\rme^{-b\mu(x)}},\qquad\qquad W(x)=-\frac{k_0b\,\rme^{-b\mu(x)}}{1-\rme^{-b\mu(x)}}+k_1,\nonumber\\
\fl\V(x)&=\frac{b^2}{2}\frac{J_\kappa(J_\kappa-1)\,\rme^{-b\mu(x)}}{\left(1-\rme^{-b\mu(x)}\right)^2}
-\frac{b^2}{2}\frac{\lam\,\rme^{-b\mu(x)}}{1-\rme^{-b\mu(x)}},\qquad\qquad
\epsilon_{0,\kappa}=-\frac{b^2}{2}\left(\frac{\lam}{2J_\kappa}-\frac{1}{2}\right)^2,\nonumber\\
\fl\xiv&=m^{1/4}(x)\exp\Bigg\{\sqrt{2}\left[\frac{\lam}{2J_\kappa}-\frac{1}{2}
-\frac{b J_\kappa\,\rme^{-b\mu(x)}}{1-\rme^{-b\mu(x)}}\right]\xi_\kappa\Bigg\}
\left(1-\rme^{-b\mu(x)}\right)^{J_\kappa}\nonumber\\
\fl&\times\exp\Bigg\{-b\left(\frac{\lam}{2J_\kappa}-\frac{1}{2}\right)\mu(x)\Bigg\},\label{5.10}
\end{eqnarray}
where $J_\kappa=(\kappa+1)k_0$ and $\lam=\left[2(\kappa+1)k_1/b+1\right]J_\kappa$. The HCS of \eref{5.10} minimize the uncertainty relation $\bra=\frac{J_\kappa b^2}{4}\,\min^2{\frac{b}{2}\mu(x)}$.

\paragraph{Hulth\'en potential (class 1):}%
It is well-known that the Manning-Rosen potential reduces to the Hulth\'en potential by setting $J_\kappa=1$, i.e.; $k_0=1/(\kappa+1)$. Thus
\begin{eqnarray}
\fl\V(x)&=-\frac{Ze^2b\,\rme^{-b\mu(x)}}{1-\rme^{-b\mu(x)}},\qquad\qquad
\epsilon_{0,\kappa}=-\frac{b^2}{2}\left(\frac{Ze^2}{b}-\frac{1}{2}\right)^2,\nonumber\\
\fl\xiv&=m^{1/4}(x)\exp\Bigg\{\sqrt{2}\left[\frac{Ze^2}{b}-\frac{1}{2}
-\frac{b\,\rme^{-b\mu(x)}}{1-\rme^{-b\mu(x)}}\right]\xi_\kappa\Bigg\}\left(1-\rme^{-b\mu(x)}\right)
\nonumber\\
\fl&\times\exp\Bigg\{-b\left(\frac{Ze^2}{b}-\frac{1}{2}\right)\mu(x)\Bigg\},\label{5.11}
\end{eqnarray}
where $Ze^2=b\lam/2$. The HCS of \eref{5.11} minimize the uncertainty relation $\bra=\frac{b^2}{4}\,\min^2{\frac{b}{2}\mu(x)}$. We note also that it is possible to recover the Yukawa potential and the accompanying HCS by setting $b\rightarrow 0$.

\paragraph{Radial harmonic oscillator potential (class 2): $(a=-1,\,b=0)$}%
\begin{eqnarray}
\fl\varphi(x)&=\frac{1}{\mu(x)},\qquad\qquad W(x)=-\frac{k_0}{\mu(x)}+k_1\mu(x)\nonumber\\
\fl\V(x)&=\frac{\om^2}{2}\,\mu^2(x)+\frac{l_\kappa(l_\kappa+1)}{2\mu^2(x)},\qquad\qquad
\epsilon_{0,\kappa}=\om\left(l_\kappa+\frac{3}{2}\right),\nonumber\\
\fl\xiv&=m^{1/4}(x)\exp\Bigg\{\sqrt{2}\left[-\frac{l_\kappa+1}{\mu(x)}+\om\mu(x)\right]\xi_\kappa\Bigg\}
\left[\mu(x)\right]^{l_\kappa+1}\exp\Bigg\{-\frac{\om}{2}\,\mu^2(x)\Bigg\},\label{5.12}
\end{eqnarray}
where $l_\kappa=(\kappa+1)k_0-1$ and $\om=(\kappa+1)k_1$. The HCS of \eref{5.12} are eigenstates of $\etat$ which minimize the uncertainty relation for $\bra=\om+\frac{l_\kappa+1}{\mu^2(x)}$.

\paragraph{Generalized P\"{o}schl-Teller potential (class 2): $(a=-b)$}%
\begin{eqnarray}
\fl\varphi(x)&=\tanh{a\mu(x)},\qquad\qquad W(x)=k_0\tanh{a\mu(x)}+k_1\coth{a\mu(x)},\nonumber\\
\fl\V(x)&=-\frac{a^2}{2}\left[\left(m_\kappa+\lam\right)^2-\frac{1}{4}\right]\max^{2}a\mu(x)
+\frac{a^2}{2}\left[\left(m_\kappa-\lam\right)^2-\frac{1}{4}\right]\min^{2}a\mu(x),\nonumber\\
\fl\epsilon_{0,\kappa}&=-\frac{a^2}{2}\left(2m_\kappa-1\right)^2,\nonumber\\
\fl\xiv&=m^{1/4}(x)\exp\Big\{\sqrt{2}\,a\left[\Lambda^{(+)}_{\kappa}\tanh a\mu(x)+\Lambda^{(-)}_{\kappa}\coth a\mu(x)\right]\xi_\kappa\Big\}\nonumber\\
\fl&\times\left[\cosh a\mu(x)\right]^{-\Lambda^{(+)}_{\kappa}}
\left[\sinh a\mu(x)\right]^{-\Lambda^{(-)}_{\kappa}},\label{5.13}
\end{eqnarray}
where $\Lambda^{(\pm)}_\kappa=m_\kappa\pm\lam-1/2$, while $m_\kappa=(\kappa+1)(k_0+k_1)/(2a)+1/2$ and $\lam=(\kappa+1)(k_0-k_1)/(2a)$. The HCS of \eref{5.13} minimize the uncertainty relation $\bra=a^2\Lambda^{(+)}_\kappa\,\max^2 a\mu(x)-a^2\Lambda^{(-)}_\kappa\,\min^2 a\mu(x)$.

\paragraph{Scarf potential (class 3): $(a=b=d=1,\,c=0)$}%
\begin{eqnarray}
\fl\varphi(x)&=\sinh{a\mu(x)},\qquad\qquad W(x)=\frac{k_0}{a}\tanh{a\mu(x)}-\frac{k_1}{a}\,\max\,a\mu(x),\nonumber\\
\fl\V(x)&=\frac{a^2}{2}\left(\lam^2-\n^2-\n\right)\max^{2}a\mu(x)
-\frac{a^2}{2}\lam\left(2\n+1\right)\tanh a\mu(x)\,\max\,a\mu(x),\nonumber\\
\fl\epsilon_{0,\kappa}&=-\frac{a^2}{2}\n^2,\nonumber\\
\fl\xiv&=m^{1/4}(x)\exp\Big\{\sqrt{2}\,a\left[\n\tanh a\mu(x)-\lam\,\max\,a\mu(x)\right]\xi_\kappa\Big\}
\left[\cosh a\mu(x)\right]^{-\n}\nonumber\\
\fl&\times\exp\Big\{-\lam\arctan\sinh a\mu(x)\Big\},\label{5.14}
\end{eqnarray}
where $\lam=(\kappa+1)k_1/a^2$ and $\n=(\kappa+1)k_0/a^2$. The HCS of \eref{5.14} are eigenstates of $\etat$ which minimize the uncertainty relation for $\bra=\frac{a^2}{2}\,\n\,\max^2 a\mu(x)+\frac{a^2}{2}\,\lam\tanh a\mu(x)\,\max\,a\mu(x)$.

\section{Conclusions}%

The main aim of this paper was to investigate the pseudo-Hermitian coherent states as well as its Hermitian counterpart coherent states in the context of the factorization method under the generalized quantum condition in the framework of position-dependent mass. We have indicated the difficulty these factorizations pose mainly due to the feature that the ladder operators do not give a closed algebra. However considering a specific modification in the superpotential instead circumvents these difficulties and, as a result, a new Hermitian counterpart Hamiltonian is deduced and factorized as a product of two adjoint ladder operators which are nothing but annihilation and creation operators for a system. As a consequence the residual algebra is nothing but the generally deformed oscillator algebra.\\
\indent Considering these ladder operators we were able to establish a general scheme to construct a wide kind of exactly solvable potentials, the associated ground-state energy and a simultaneous derivation of their corresponding HCS is then possible under a generalized quantum condition term $U^2(x)W'(x)=F[x]$. It is found that this term is rather simple to deduce by making use of \eref{5.1}, \eref{5.2} and \eref{5.3}. Alternatively by imposing a similarity transformation between the pseudo-Hermitian Hamiltonian and its Hermitian counterpart we have constructed their accompanying PHCS.\\
\indent We have shown that HCS and PHCS are the same analytical form as Perelomov's states and thus minimize a generalized position-momentum uncertainty principal.

\section*{References}%

\end{document}